\documentclass{emulateapj}
\submitted{Accepted in ApJ, 2007 December 2}

\newcommand\psr{1E~1048.1$-$5937}
\newcommand\tfn{1E~2259$+$586}

\newcommand\hsth{$m_{F160W}$}
\newcommand\hstj{$m_{F110W}$}

\newcommand\xte{\textit{RXTE}}
\newcommand\cxo{\textit{CXO}}
\newcommand\xmm{\textit{XMM}}
\newcommand\swift{\textit{Swift}}
\newcommand\hst{\textit{HST}}
\newcommand\vlt{\textit{VLT}}

\defcitealias{gk04}{GK04}

\begin{document}

\title{X-ray and Near-IR Variability of the Anomalous X-ray Pulsar
  \psr: From Quiescence Back to Activity}

\author{
  Cindy R. Tam,\altaffilmark{1}
  Fotis P. Gavriil,\altaffilmark{2,3}
  Rim Dib,\altaffilmark{1}
  Victoria M. Kaspi,\altaffilmark{1,4}
  Peter M. Woods, \altaffilmark{5,6}
  Cees Bassa,\altaffilmark{1}
}

\altaffiltext{1}{Department of Physics, Rutherford Physics Building,
  McGill University, 3600 University Street, Montreal, QC, H3A 2T8,
  Canada} \altaffiltext{2}{NASA Goddard Space Flight Center,
  Astrophysics Science Division, Code 662, Greenbelt, MD, 20771, USA}
\altaffiltext{3}{NPP Fellow; Oak Ridge Associated Universities,
  Building SC$-$200, 1299 Bethel Valley Road, Oak Ridge, TN, 37830, USA}
\altaffiltext{4}{Canada Research Chair; Lorne Trottier Chair;
  R. Howard Webster Fellow of CIFAR} \altaffiltext{5}{Dynetics, Inc.,
  1000 Explorer Boulevard, Huntsville, AL, 35806, USA}
\altaffiltext{6}{NSSTC, 320 Sparkman Drive, Huntsville, AL, 35805,
  USA}

\begin{abstract}
  We report on new and archival X-ray and near-infrared (near-IR)
  observations of the anomalous X-ray pulsar \psr\ performed between
  2001-2007 with the \textit{Rossi X-ray Timing Explorer} (\xte), the
  \textit{Chandra X-ray Observatory} (\cxo), the \textit{Swift
  Gamma-ray Burst Explorer}, the \textit{Hubble Space Telescope}
  (\hst), and the \textit{Very Large Telescope}.  During
  its $\sim$2001-2004 active period, \psr\ exhibited two large,
  long-term X-ray pulsed-flux flares as well as short bursts, and
  large ($>$10$\times$) torque changes.  Monitoring with \xte\
  revealed that the source entered a phase of timing stability in
  2004; at the same time, a series of four simultaneous observations
  with \cxo\ and \hst\ in 2006 showed that its X-ray flux and spectrum
  and near-IR flux, all variable prior to 2005, stabilized.
  Specifically, we find the 2006 X-ray spectrum to be consistent with 
  a two-component blackbody plus power law, with average $kT=0.52$~keV
  and power-law index $\Gamma=2.8$ at a mean flux level in the
  2$-$10~keV range of $\sim$6.5$\times
  10^{-12}$~erg~cm$^{-2}$~s$^{-1}$.  The near-IR flux, when detected
  by \hst\ ($H\sim22.7$~mag) and \vlt\ ($K_S\sim21.0$~ mag), was
  considerably fainter than previously measured.  Recently, in 2007
  March, this newfound quiescence was interrupted by a sudden flux
  enhancement, X-ray spectral changes and a pulse morphology change,
  simultaneous with a large spin-up glitch and near-IR enhancement.
  Our \xte\ observations revealed a sudden pulsed flux increase by a
  factor of $\sim$3 in the 2$-$10~keV band.  In observations with
  \cxo\ and \swift, we found that the total X-ray flux increased much
  more than the pulsed flux, reaching a peak value of $>$7 times the
  quiescent value (2$-$10~keV).  With these recent data, we
  find a strong anti-correlation between X-ray flux and pulsed
  fraction. In addition, we find a correlation between X-ray spectral
  hardness and flux.  Simultaneously with the radiative and timing
  changes, we observed a significant X-ray pulse morphology change
  such that the profile went from nearly sinusoidal to having multiple
  peaks.  We compare these remarkable events with other AXP outbursts
  and discuss implications in the context of the magnetar model and
  other models of AXP emission.
\end{abstract}


\section{Introduction}

The small handful of unusual objects known as anomalous X-ray
pulsars (AXPs) are characterized by their slow spin periods (5--12~s),
large inferred magnetic fields ($\sim$10$^{14}$~G), and persistent
X-ray luminosities in great excess of available spin-down power.
These young, isolated pulsars have properties common to another
class of objects, the soft gamma repeaters (SGRs).  For recent
reviews, see \citet{wt06} and \citet{kas07}.  Potentially the most
intriguing and most revealing property shared by
AXPs and SGRs is their highly volatile nature: both are known to
exhibit occasional and sudden dramatic flux and spin variability in
the form of X-ray bursts, flares, and glitches.

The magnetar model \citep{dt92,td95,td96a} was instrumental in
explaining both the ``anomalous'' X-ray emission, as well the episodes
of burst activity.  It identified the sources as highly magnetized
isolated neutron stars, with field strengths on the order of
$10^{14}-10^{15}$~G.  The magnetar model uniquely, and correctly,
predicted that SGR-like bursts would be observed from AXPs. A major
proposed effect of this enormous magnetization is the ``twisting'' of
the magnetosphere, with associated magnetospheric currents and heating
of the crust at the twisted field-line footpoints \citep{tlk02}.  This
gives rise to the unusual X-ray spectrum, which has generally been
empirically well characterized by a two-component model consisting of
a power law plus blackbody. Bursts of high-energy emission occur
presumably when the crust succumbs to magnetic stresses and deforms,
leading to a rapid rearrangement of the external magnetic field.
Recently, \citet{bt07} proposed the existence of a plasma corona
contained within the closed magnetosphere to explain the broad band
spectrum of magnetars that extends from the infrared (IR) to hard
X-rays beyond 100~keV \citep{khdc06}.

\psr\ has had an unusual history, even by AXP standards\footnote{A
summary of its properties can be found in the online SGR/AXP
Catalog
\texttt{http://www.physics.mcgill.ca/$\sim$pulsar/magnetar/main.html}.}. 
Prior to 2002, monitoring of the pulsed flux with the
\textit{Rossi X-ray Timing Explorer} (\textit{RXTE}) showed that
its spin down was so unstable that phase coherence could be maintained
for periods of only months at a time \citep{kgc+01}, behaviour which
echoed earlier reports of $\dot{P}$, flux and spectral variability
\citep{opmi98}. Two small bursts, the first ever found in an AXP, were
discovered in 2001 from \psr\ \citep{gkw02}, clinching the suspected
association between AXPs and SGRs.  Since 2002, tri-weekly
\textit{RXTE} observations of \psr\ allowed the changing pulsed flux
and spin-down rate to be monitored on relatively short timescales.
\citet[][hereafter GK04]{gk04} reported order-of-magnitude
torque variability on timescales of weeks to months that only
marginally correlated with large luminosity variations.  The unusually
slow and long-lived pulsed flux ``flares'' (not to be confused with
the SGR giant flares; see \citealp{wt06}) that began in late 2001 and
lasted into 2004 are in some ways unlike activity seen in any other
AXP or SGR thus far. The 2$-$10~keV spectrum and pulsed fraction are
also variable on long timescales, the latter being anti-correlated
with luminosity \citep{mts+04,tmt+05}.  At near-infrared (near-IR)
wavelengths, the flux varied dramatically during 2002$-$2003
\citep{wc02,ics+02}; unfortunately, the sparsity of near-IR
observations prohibits measurements of the variability timescale,
although a near-IR/X-ray anti-correlation has been suggested
\citep{dv05a}.  Further activity in the form of another burst and
small pulsed-flux increase was seen in 2004 \citep{gkw06}.

Here, we report on a series of simultaneous \cxo\ and \hst\
observations obtained during the course of 2006, as well as on
archival \cxo, \xmm\ and \vlt\ observations, in
\S~\ref{sec:observations}. We also present \xte\ timing results from
the past three years.  In 2007 March, \psr\ unexpectedly entered
yet another new phase of activity, and we report on
Target-of-Opportunity observations with \cxo\ and \swift\ taken after
the 2007 March event.  In \S~\ref{sec:quiescent 
discussion} we discuss the 2004-2006 period of radiative and spin
quiescence in the context of both magentar and accretion models, and
in \S~\ref{sec:flare discussion}, we describe and interpret the events
following the 2007 March flare.

\section{Observations, Analysis and Results}
\label{sec:observations}

As part of a long-term project, we regularly monitor AXPs using the
\textit{Rossi X-ray Timing Explorer} (\xte). \psr\ is by far the most
frequently observed AXP due to its relatively poor rotational
stability. The \xte\ observations (see \S~\ref{sec:RXTE}) are crucial
in measuring the spin evolution of the source and its pulsed
flux. However, the high background and large unimaged field-of-view of
\xte\ make spectral measurements of \psr\ difficult and total flux
(and consequentially pulsed fraction) measurements
impossible. To make these measurements, we obtained five
\textit{Chandra X-ray Observatory} observations (\cxo; see
\S~\ref{sec:Chandra}) approximately equispaced throughout 2006.  In
order to probe the origin of the near-IR emission of this source, we
observed with the \textit{Hubble Space Telescope} (\hst; see
\S~\ref{sec:hst}) simultaneously with \textit{CXO}. The motivation for
simultaneous \cxo\ and \hst\ observations was to search for correlated
variability in the different bands, as that provides insight into the
physical mechanisms generating the radiation.  We also examined
archival $K_S$-band observations with the \textit{Very Large
Telescope} (\vlt; see \S~\ref{sec:vlt}), which contribute to our
monitoring of its near-IR brightness. Following the 2007 March glitch
detected in our \xte\ monitoring observations of \psr\ \citep{dkgw07},
we initiated three \cxo\ and two \textit{Swift Gamma-Ray Burst
Explorer} (\swift; see \S~\ref{sec:Swift}) Target-of-Opportunity (ToO)
observations to follow the X-ray flux and spectrum of the source.  In
order to characterize the long-term evolution of \psr\ we also
incorporate and reanalyze archival \textit{XMM-Newton} and \cxo\
observations \citep[see][for details on how these data were
processed]{gkw06}.  Separate post-outburst optical and near-IR
observations of \psr\ are reported in detail elsewhere
\citep{wbk+07}.

\subsection{\xte}
\label{sec:RXTE}

\begin{figure}
%\epsscale{0.75} %%% MANUSCRIPT
\plotone{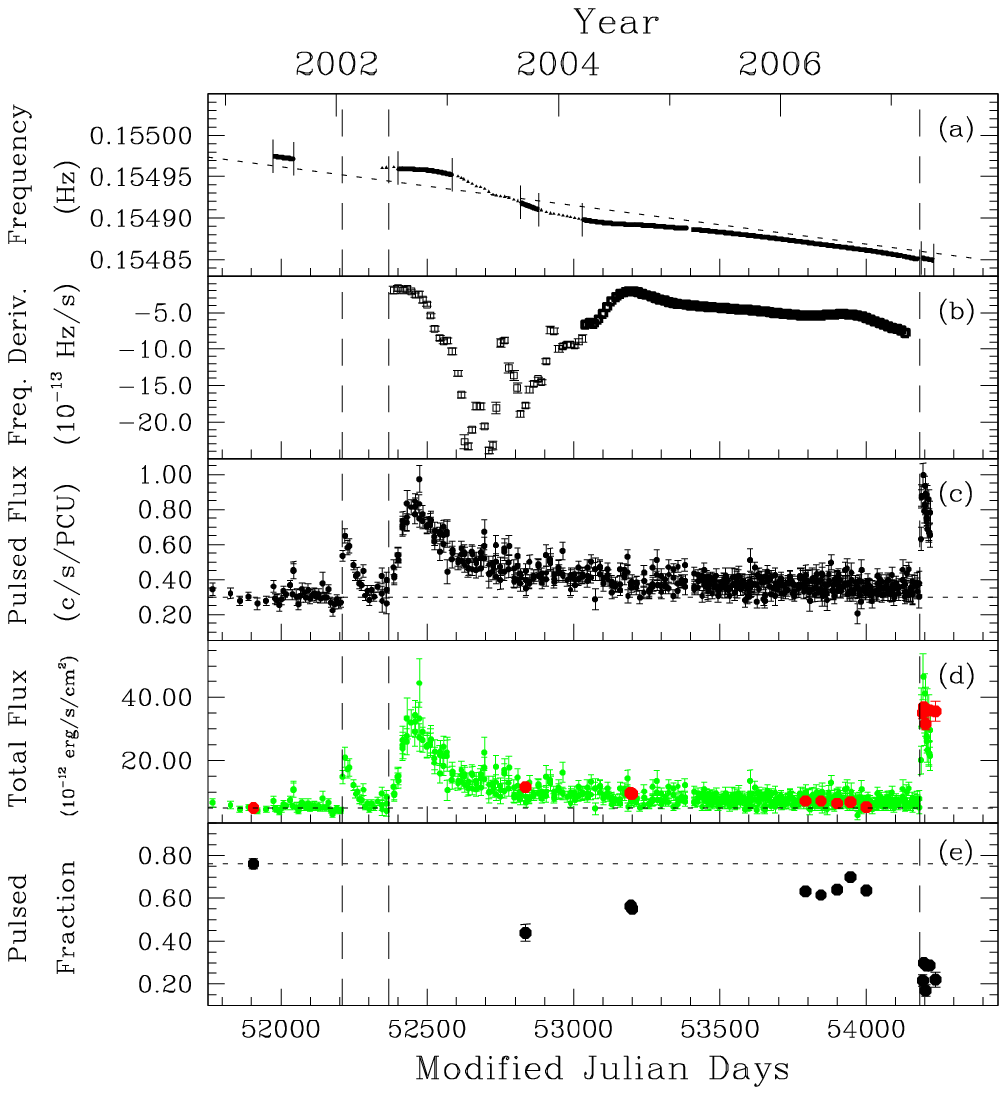}
\figcaption{The long-term evolution of \psr's pulsed properties.
  The three vertical dashed lines indicate the approximate beginnings
  of the flux flares on (from left to right) 2001 October 26, 2002
  April 6, and 2007 March 21, respectively.  Horizontal dotted lines
  represent the value of the plotted parameter prior to all observed
  flares. 
  (\textit{a}) Spin frequency as observed by \xte.  Small points
  show the measured frequencies, while intervals over which phase
  coherence has been maintained are shown as thick lines
  (\citetalias{gk04}; Dib et al., in preparation).
  (\textit{b}) Frequency derivative as a function of time.  While the
  frequency was varying in 2002-2004 and a phase-coherent timing
  solution could not be found, $\dot{\nu}$ was determined in short
  intervals by calculating the slope of three consecutive $\nu$
  measurements \citepalias[after][]{gk04}.  Since 2004, phase
  coherence has been maintained and $\dot{\nu}$ reflects the standard
  technique (Dib et al., in preparation). 
  (\textit{c}) 2$-$10~keV rms pulsed flux as observed by \xte.
  (\textit{d}) Simulated total 2$-$10~keV unabsorbed flux, shown as
  green points.  Total flux is estimated from the \xte\ pulsed flux
  and the power-law correlation between pulsed fraction and measured 
  total flux, shown as red points, as described in \S~\ref{sec:pfvsf
  disc}.  The simulated fluxes (\textit{green}) have been scaled to
  match the actual measured total fluxes (\textit{red}). 
  (\textit{e}) 2$-$10~keV rms pulsed fraction calculated using the
  method described in \citet{wkt+04}. 
  \label{fig:timing}}
\end{figure}

We have observed \psr\ regularly since 1997 with \xte\
(\citealt{kgc+01}; \citetalias{gk04}). Our data were obtained using
the Proportional Counter Array (PCA) on board \xte\ which consists of
five identical and independent Xenon/Methane Proportional Counter
Units (PCUs).  We use our \xte\ observations of \psr\ to look for the
presence of bursts (see \citealt{gkw04} for details), to look for
pulse profile changes, to monitor its pulsed flux, and to monitor its
frequency evolution using phase-coherent timing when possible.

For the timing analysis, we created barycentered lightcurves in the
2$-$5.5~keV band with 31.25~ms time resolution.  As past monitoring
has shown that it can be difficult to maintain pulse phase coherence
over timescales longer than a few weeks, observations of \psr\ since
2002 are done three times per week, with the three observations
carefully spaced so as to allow a phase-coherent analysis and a
precise frequency measurement.  Thus, for each observation, we fold at
the pulse period determined via periodogram, cross-correlate the
folded profiles with a high signal-to-noise template, and fit the
resulting phases with a linear function whose slope provides the
average frequency.  Frequencies determined in this way are shown in 
Figure~\ref{fig:timing}.
As can be seen in this figure, the 2004-2006 frequencies were
actually much more stable than in the past; we therefore attempted a
fully phase-coherent analysis as well.  This will be described
elsewhere (Dib et al., in preparation).  Frequency derivatives,
which are also displayed in Figure~\ref{fig:timing} and are from
\citetalias{gk04} and Dib et al., (in preparation), are clearly much
more stable in the 2004-2006 interval than previously.

This period of rotational stability was accompanied by X-ray pulsed
flux stability and relative quiescence:  the pulsed flux time series
of \psr\ in the 2$-$10~keV band is presented in
Figure~\ref{fig:timing}. The pulsed flux was calculated
using a method similar to that described in \citet{wkt+04} and is
based on the rms of the folded profile, but without variance
subtraction.

The long period of rotational stability ended in 2007 March, when a
large glitch was observed \citep{dkgw07}. The glitch also signaled the
end of the period of pulsed flux quiescence. The pulsed flux suddenly
increased by a factor of $\sim$3 in the energy range 2$-$10~keV.  The
upper limit on the rise time of the pulsed flux for this event is
approximately one week. The peak pulsed flux reached by the source was
$\sim$10\% larger than the peak flux reached during the largest of the
two previously observed flares, and the rise time was at least 4 times
smaller. On 2007 May 17, the date of the last X-ray imaging observation
included in this paper, the pulsed flux had decreased by $\sim$10\%. 
It is as yet difficult to compare the decay timescale of the pulsed
flux of this new event to that of the previous flares however it
will be possible in the near future.

\subsection{\cxo\ and \swift}
\label{sec:Chandra}
\label{sec:Swift}
X-ray imaging observations were carried out with the \cxo\ and \swift\
telescopes. The date, total exposure time and resulting
count rate for each region-filtered background-subtracted observation
are listed in Table~\ref{tab:obs} and a detailed description of our
spectral fitting is given in \S~\ref{sec:Spectral Fit}.

\begin{figure}
%\epsscale{0.75} %%% MANUSCRIPT
\plotone{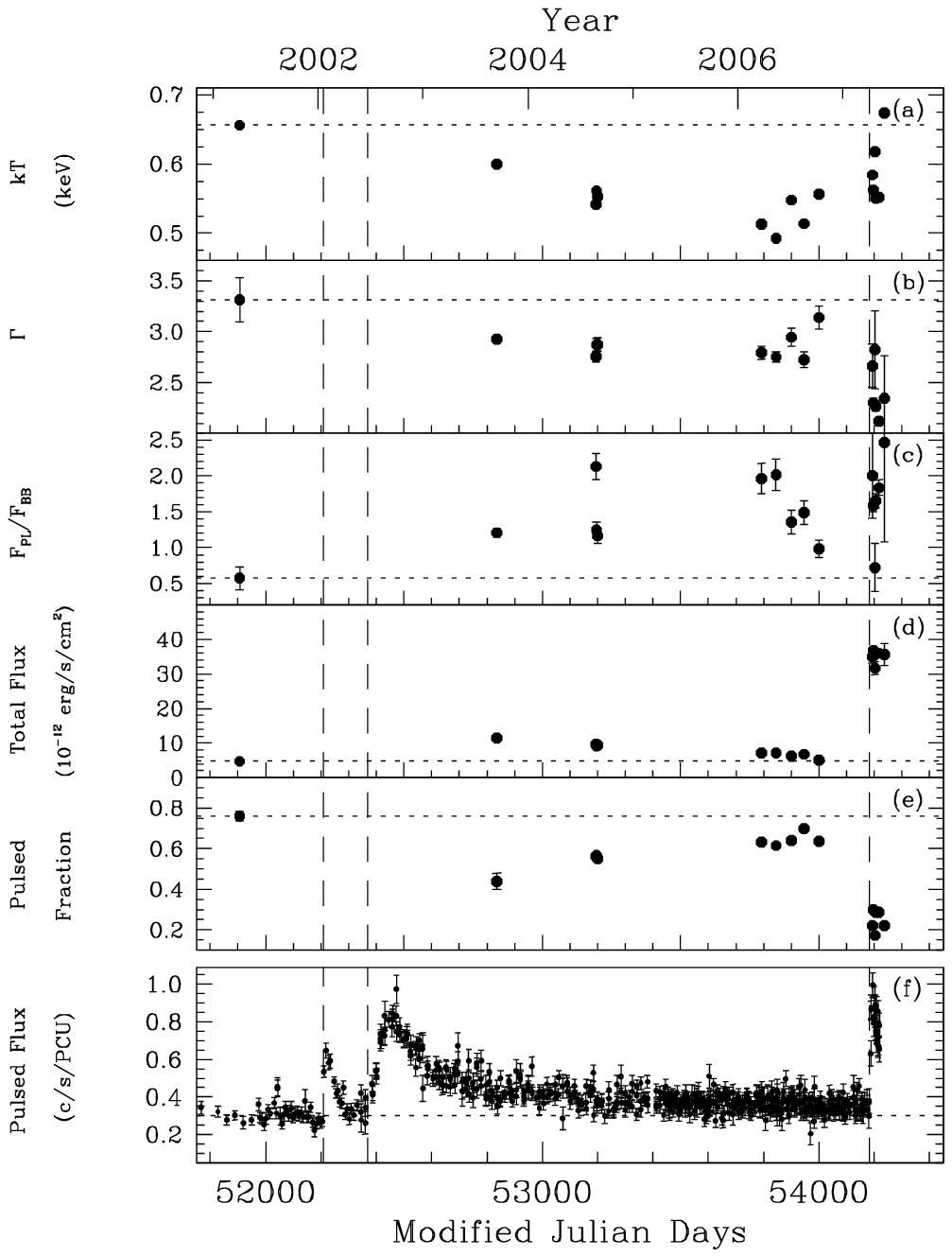}
\figcaption{The evolution of \psr's X-ray spectral properties. For
  consistency, all the spectra from \cxo, \swift, and \xmm\ were fit
  jointly with a two-component absorbed blackbody plus power-law
  model, that produced a best-fit $N_H=(0.97\pm
  0.01)\times10^{22}$~cm$^{-2}$ (see \S~\ref{sec:Spectral Fit}). The
  uncertainties shown reflect statistical errors only.
  Vertical dashed lines indicate the beginnings of the flux flares
  (see Fig.~\ref{fig:timing} caption),
  while horizontal dotted lines indicate quiescent values.  
  (\textit{a}) Blackbody temperature $kT$.
  (\textit{b}) Photon index $\Gamma$.
  (\textit{c}) Ratio of the 2$-$10~keV flux contribution from the
  blackbody and power law components.
  (\textit{d}) Total unabsorbed flux in 2$-$10~keV.
  (\textit{e}) 2$-$10~keV rms pulsed fraction calculated using the
  method described in \citet{wkt+04}.
  (\textit{f}) For reference, we show the 2$-$10~keV rms pulsed flux
  as observed by \xte. 
  \label{fig:xrayspec}}
\end{figure}

\psr\ was observed five times\footnote{Although four simultaneous
observations were originally planned, \hst\ experienced technical
difficulties during one, resulting in a fifth \cxo\ observation.} in
2006 with \cxo\ in quasi-equispaced intervals, and three times
in 2007 April.  The observations were made with the Advanced CCD
Imaging Spectrometer (ACIS) camera using the S3 chip in continuous
clocking (CC) mode. This mode generates a 1$\times$1024 pixel
image, which is read out every 2.8~ms; pile-up in this mode is not an
issue. The subsequent analysis was based on the ``Level 2''
events files for which the event times are photon arrival times (as
opposed to readout times for the ``Level 1'' events). Following the
standard threads we then corrected the ``Level 2'' events for certain
caveats related to ACIS CC-mode
data\footnote{\texttt{http://cxc.harvard.edu/ciao3.4/why/ccmode.html}}.
We extracted a rectangular region centered on the source with a width
typically of $\sim$30 pixels.  We estimated the background using
rectangular regions on either side of the source which extended from
$\sim$10 pixels beyond the edge of the source region to $\sim$10
pixels from the edge of the image.  Following the standard threads,
the source and background spectra were then extracted using
\texttt{CIAO}\footnote{\texttt{http://cxc.harvard.edu/ciao}} v3.4.  We
grouped the source spectrum such that there were no fewer than 20
counts per bin after background subtraction.  The response matrix
function (rmf) and area response function (arf) for each observation
were also generated using \texttt{CIAO} and \texttt{CALDB}. We used
Timed Exposure (TE) mode response matrices because of the absence of
available spectral calibration for CC mode.

\psr\ was observed three times\footnote{Two additional
\swift\ observations taken contemporaneously were omitted from our
analysis because their short exposures yielded prohibitively large
uncertainties.} in 2007 with \swift\ using the X-ray
Telescope (XRT). XRT has two observing modes: photon counting (PC) and
windowed-timing (WT). PC mode provides a $\sim$600$\times$600
pixel$^2$ image with low ($\sim$2.5~s) time resolution, whereas WT
mode provides a one-dimensional $\sim$200-pixel wide image with high
($\sim$0.74~ms) time resolution. XRT automatically switches from WT to
PC-mode when the count rate exceeds $\sim$2 counts s$^{-1}$. In our
analysis we ignored the WT mode data as they were too short to be of
use. We used the PC-mode event files from the standard pipeline; these
were cleaned and barycentered.  Using \texttt{FTOOL xrtcentroid}, we
determined the centroid of the image, and input the cleaned,
barycentered events into the command line interface
\texttt{xselect}\footnote{\texttt{http://heasarc.gsfc.nasa.gov/docs/software/lheasoft/ftools/xselect/xselect.html}}.
We selected a circular source region with a 25 pixel radius centered
on the centroid of the image. Since the source was slightly piled up,
we excluded a 4 pixel radius circular region centered on the centroid
of the image. We used 4 pixels because that is the recommended
exclusion radius for a source with \psr's spectrum (L. Angelini,
private communication). For the background, we selected events from an
annulus with an
inner radius of 50 pixels and outer radius of 130 pixels. We further
filtered our events by selecting only those that had grades 0 to
12. With \texttt{xselect} we then created source and background PHA
files. The \texttt{FTOOL xrtmkarf} was used to create the arfs, and we
input the source PHA file with the excluded center so that it would be
corrected for pile-up. We used the file provided by \texttt{CALDB}
appropriate for PC mode data events of grades 0 to 12 to make the rmf.

%\begin{deluxetable}{lccccccccc} %%% MANUSCRIPT
\begin{deluxetable*}{lccccccccc} %%% EMULATEAPJ
%\tabletypesize{\scriptsize} %%% MANUSCRIPT
%\rotate                   %%% MANUSCRIPT
%\tablewidth{0pt}          %%% MANUSCRIPT
\tablewidth{2\columnwidth} %%% EMULATEAPJ
\tablecaption{\cxo, \xmm\ and \swift\ observing parameters and results
  \label{tab:obs}}
\tablehead{\colhead{Date} & \colhead{MJD} & \colhead{Obs. ID} & \colhead{Exposure} &
  \colhead{Count rate} & \colhead{$\Gamma$\tablenotemark{a}}    &
  \colhead{$kT$\tablenotemark{a}} &
  \colhead{Unabs. Flux\tablenotemark{a,b}} &
  \colhead{$F_{\mathrm{PL}}/F_{\mathrm{BB}}$\tablenotemark{a,c}}
  &\colhead{Pulsed}     \\
  \colhead{} & \colhead{} & \colhead{} & \colhead{(s)} & \colhead{(counts
    s$^{-1}$)} & \colhead{} & \colhead{(keV)} & \colhead{(10$^{-12}$
    erg cm$^{-2}$ s$^{-1}$)} &\colhead{}  &\colhead{
    Fraction\tablenotemark{d}}   } 
\startdata  
\multicolumn{10}{c}{Archival  \xmm\ Observations}\\
\tableline
2000 Dec 28 & 51906 & 0112780401 & 4447  &  1.461(18) & 3.31(22) & 0.656(17)  &  4.8(4)  & 0.6(2)  & 0.76(2)  \\
2003 Jul 16 & 52836 & 0147860101 & 41145 &  3.429(9)  & 2.93(4)  & 0.599(6)  &  11.5(1)  & 1.21(5) & 0.44(4)  \\
2004 Jul 9  & 53195 & 0164570301 & 32679 &  0.985(6)  & 2.75(5)  & 0.542(15)  &  9.6(1)  & 2.1(2)  & 0.561(7) \\
\tableline
\multicolumn{10}{c}{Archival \cxo\ Observations} \\
\tableline
2004 Jul 10 & 53196 & 4653  & 28860 & 1.363(7) & 2.87(7) & 0.56(1)  & 9.3(1) & 1.2(1) & 0.56(1) \\
2004 Jul 15 & 53201 & 4654  & 28085 & 1.317(7) & 2.87(7) & 0.55(1)  & 9.4(2) & 1.2(1) & 0.55(1) \\
\tableline
\multicolumn{10}{c}{Monitoring \cxo\ Observations} \\
\tableline
2006 Feb 26 & 53792 & 6733  & 22085 & 1.072(8) & 2.79(6) & 0.51(1)  & 7.1(1)  & 2.0(2) & 0.63(1) \\
2006 Apr 20 & 53845 & 6734  & 20553 & 1.157(9) & 2.75(5) & 0.50(1)  & 7.1(2)  & 2.0(2)  & 0.61(1) \\
2006 Jun 14 & 53900 & 6735  & 22085 & 0.985(9) & 2.95(9) & 0.55(1)  & 6.3(1)  & 1.4(2)  & 0.64(1) \\
2006 Jul 30 & 53946 & 7347  & 22176 & 0.821(8) & 2.72(8) & 0.51(1)  & 6.8(2)  & 1.2(2)  & 0.70(1) \\
2006 Sep 23 & 54001 & 6736  & 22022 & 1.006(7) & 3.14(11) & 0.56(1) & 5.1(1)  & 1.0(2) & 0.64(1) \\
\tableline
\multicolumn{10}{c}{ToO \cxo\ Observations} \\
\tableline
2007 Apr 6  & 54196 & 7647 & 20079 & 4.241(15) & 2.30(4) & 0.56(1)   & 36.9(5) & 1.6(1) & 0.299(5) \\
2007 Apr 16 & 54206 & 7648 & 18387 & 3.959(15) & 2.27(4) & 0.55(1)   & 36.0(5) & 1.7(1) & 0.286(6) \\
2007 Apr 28 & 54218 & 7649 & 19082 & 4.332(15) & 2.12(4) & 0.55(1)   & 36.0(5) & 1.8(1) & 0.287(5) \\
\tableline
\multicolumn{10}{c}{ToO \swift\ Observations} \\
\tableline
2007 Apr 3  & 54193 &  30912001 & 4905  &  0.58(1)  & 2.7(2)   & 0.58(6) & 35(2)  & 2.0(6) & 0.22(3)\\
2007 Apr 13 & 54203 &  30912003 & 4829  &  0.55(1)  & 2.8(4)   & 0.62(4) & 32(2)  & 0.7(3) & 0.17(3)\\
2007 May 17 & 54237 &  30912009 & 3626  &  0.51(1)  & 2.3(4)   & 0.7(1) & 35(3)   & 2(1)   & 0.22(3)
\enddata
\tablecomments{All quoted errors represent 1$\sigma$
  uncertainties. The archival \xmm\ and \cxo\ observations have been
  previously reported by \citet{mts+04}, \citet{tmt+05}, and
  \citet{gkw06}.}
\tablenotetext{a}{These results are from a simultaneous spectral fit
  to all the data using as a model a photoelectrically absorbed
  blackbody with temperature $kT$ and a power law with photon index
  $\Gamma$. In the simultaneous fit the column density, $N_H$, was
  constrained to be the same for all observations. The resultant $N_H$
  from the global fit was $N_H=(0.97\pm 0.01)\times10^{22}$~cm$^{-2}$;
  see the text for details.}
\tablenotetext{b}{Total unabsorbed flux in the 2$-$10~keV band.}
\tablenotetext{c}{Ratio of the 2$-$10 keV flux in the power-law 
  component, $F_{\mathrm{PL}}$, to the 2$-$10 keV flux in the
  blackbody component, $F_{\mathrm{BB}}$.}  
\tablenotetext{d}{2$-$10 keV pulsed fraction, using the definition of
  pulsed fraction described in \citet{wkt+04}.} 
%\end{deluxetable} %%% MANUSCRIPT
\end{deluxetable*} %%% EMULATEAPJ

\subsubsection{Global Spectral Fit}
\label{sec:Spectral Fit}
We extracted spectra of \psr\ from all archival, monitoring, and ToO
observations collected from \cxo, \xmm\ and \swift.  A description of
the recent data analysis is in \S~\ref{sec:Chandra}, and the archival
analysis was described in \citet{gkw06}.  Using the fitting package
\texttt{XSPEC}\footnote{\texttt{http://xspec.gsfc.nasa.gov}} v12.3.1
we modeled the X-ray spectra with a photoelectrically absorbed
blackbody plus power law.  We used the \texttt{phabs} photoelectric
absorption \texttt{XSPEC} model, which assumes the solar abundance
table of \citet{ag89} and uses the \texttt{bcmc} photoionization
cross-section table from \citet{bm92} with the new He cross-section
from \citet{bm98}.  We fit this model to all the observations
simultaneously, allowing the column density $N_H$ to vary but with the
sole constraint that it be the same for all observations. We
restricted our fit to the 0.7--5~keV band. The total global fit had
$\chi^2_{\nu}=1.07$ for $\nu=4200$ degrees of freedom (dof).
Individually, the observations were equally well modeled (except for a
relatively high $\chi^2$ value in one observation, see
\S~\ref{sec:spec feature}). Our best-fit column density obtained from
the global fit is $N_H=(0.97\pm0.01)\times10^{22}$~cm$^{-2}$, in
agreement with \citet{dv06b}. The results of our spectral fitting are
listed in Table~\ref{tab:obs} and plotted in
Figure~\ref{fig:xrayspec}.

\subsubsection{Possible Spectral Feature}
\label{sec:spec feature}
\begin{figure}
%\epsscale{0.75} %%% MANUSCRIPT
\plotone{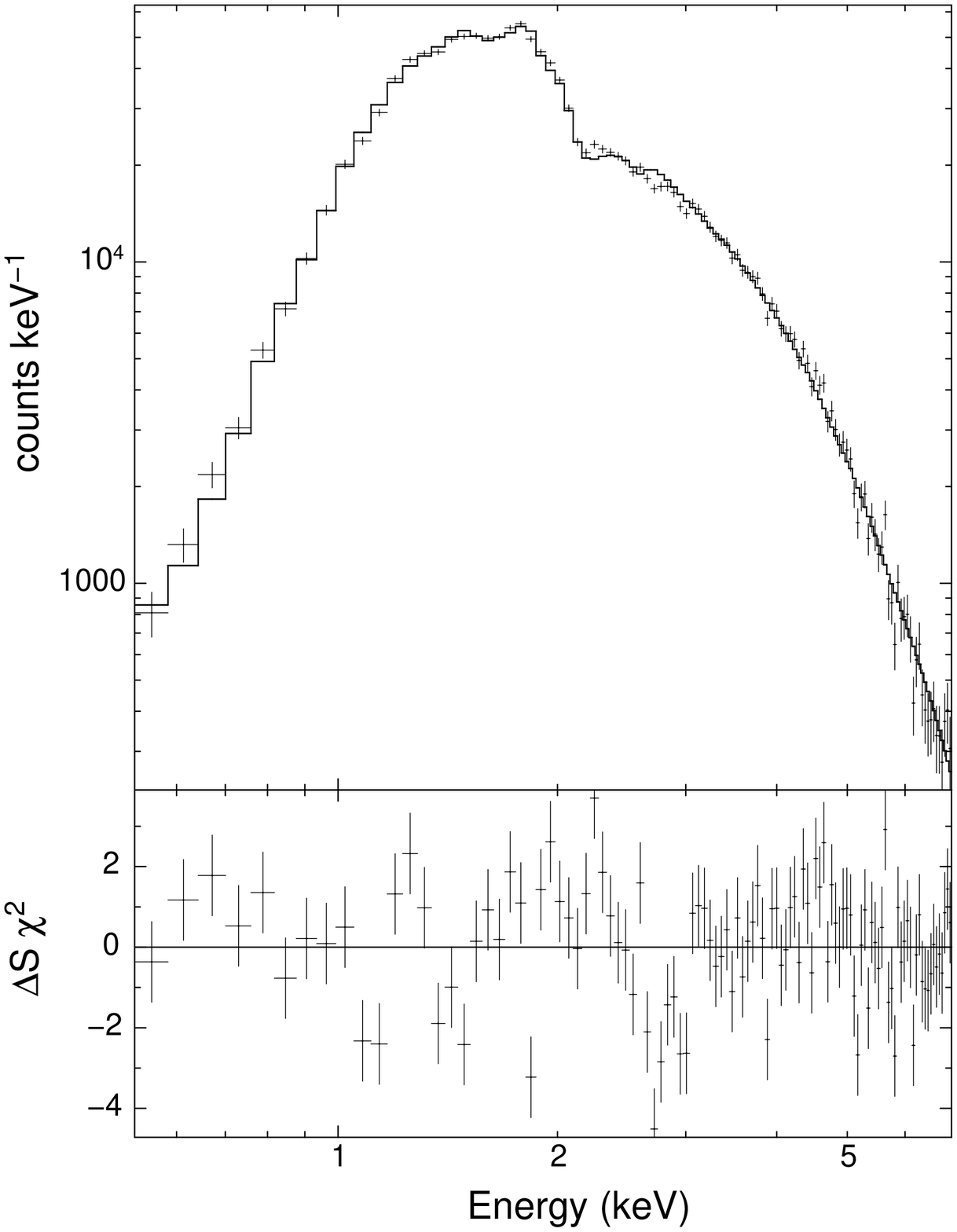} 
\figcaption{\psr's X-ray spectrum on 2007
  April 6, as observed by {\it CXO}. \textit{Top}: The best-fit
  photoelectrically absorbed blackbody plus power law is plotted as a
  solid line; see text for details on the analysis. The quality of the
  fit was low, with $\chi^2_{\nu}=1.33$ for $\nu=353$
  dof. \textit{Bottom}: Residuals after subtracting the best-fit
  model.  Notice a possible spectral feature at
  $\sim$2.7~keV. \label{fig:spectrum}}
\end{figure}

All observations had spectra that were well modeled by a
photoelectrically absorbed blackbody plus power law.  However, the
\cxo\ observation on 2007 April 6 (observation ID 7647), the first
after the 2007 March event, had a relatively high $\chi^2_{\nu} =
1.33$ ($\nu=353$ dof). There is possible evidence for an absorption
line at $\sim$2.7~keV (see Fig.~\ref{fig:spectrum}).  Adding a
Gaussian line improved the fit ($\chi^2_{\nu}=1.20$ for
$\nu=351$ dof) with $\Delta \chi^2=48.3$.  With the line width fixed
at 0.1~keV, we measure a line energy of $2.73\pm0.03$~keV. Allowing
the line width $\sigma$ to be a free parameter, we obtain
$\sigma=0.16\pm0.03$~keV, a line energy of $2.74\pm0.03$~keV and
$\chi^2_{\nu} = 1.19$ for 351 dof.  We find no evidence of a
phase dependence for this possible line.  A detailed phase-resolved
spectroscopic analysis will be presented in a forthcoming paper.

To test the significance of
adding such a line, we performed the following simulation.  We
generated 10000 simulated spectra by adding Poisson noise to a
blackbody plus power-law model spectrum having the same parameters as
our best-fit model.  We then determined the maximum change in $\chi^2$
after adding a Gaussian line. To avoid local minima and ensure that
we found the true minimum $\chi^2$ for each simulation iteration,
instead of fitting for the peak energy, we stepped through different
line energies between 0.6 and 7.0~keV with a step size of 0.025~keV.
The width of each line was held fixed at 0.1~keV and its normalization
was allowed to vary. In all the simulation iterations, none had a
change in $\chi^2$ greater than 48.3. The probability that
this feature is due to random chance is $<$0.13\%, accounting for the
number of trials.  We note however that with the addition of the line,
the overall fit is still unacceptable; we speculate that other
features may be present as well, though at marginal significance.  For
example, there is a small but intriguing feature at $\sim$1.3~keV,
which is approximately half the energy of the line discussed above.
The evidence for a line must be considered tentative because
CC mode, which uses the TE mode response matrices, is not spectrally
calibrated, and there exist calibration lines between 1.7 and 3~keV in
the TE response. However, this is true for the other \cxo\ observations,
including those in which we find no such features.

\subsubsection{Pulse Morphology and Pulsed Fraction Study}
\label{sec:profile and fraction}
Using our \cxo\ and \swift\ observations we were also able to
study the source's pulse morphology, flux and pulsed fraction.

\begin{figure*} %%% EMULATEAPJ
%\epsscale{1.0} %%% MANUSCRIPT
\plottwo{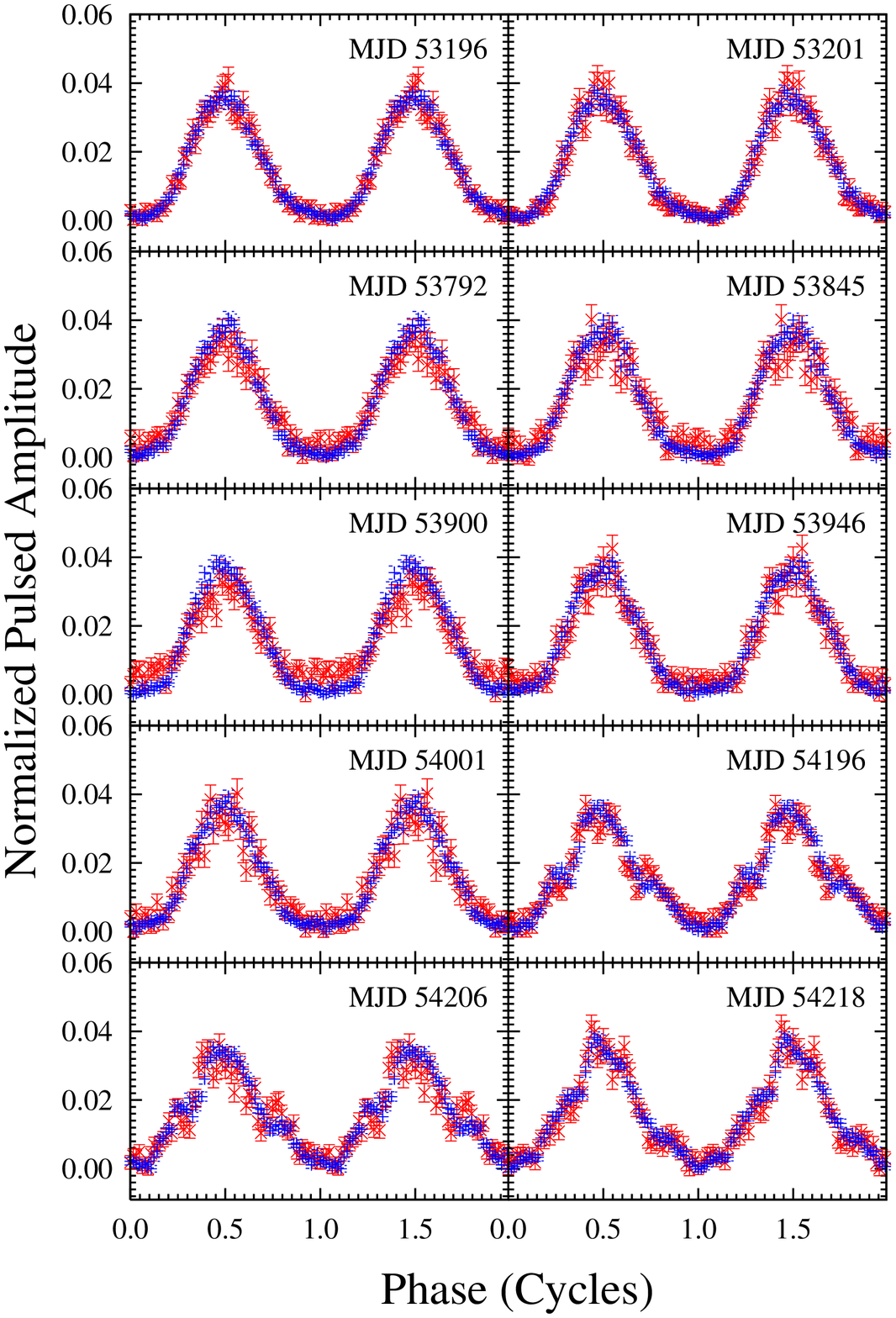}{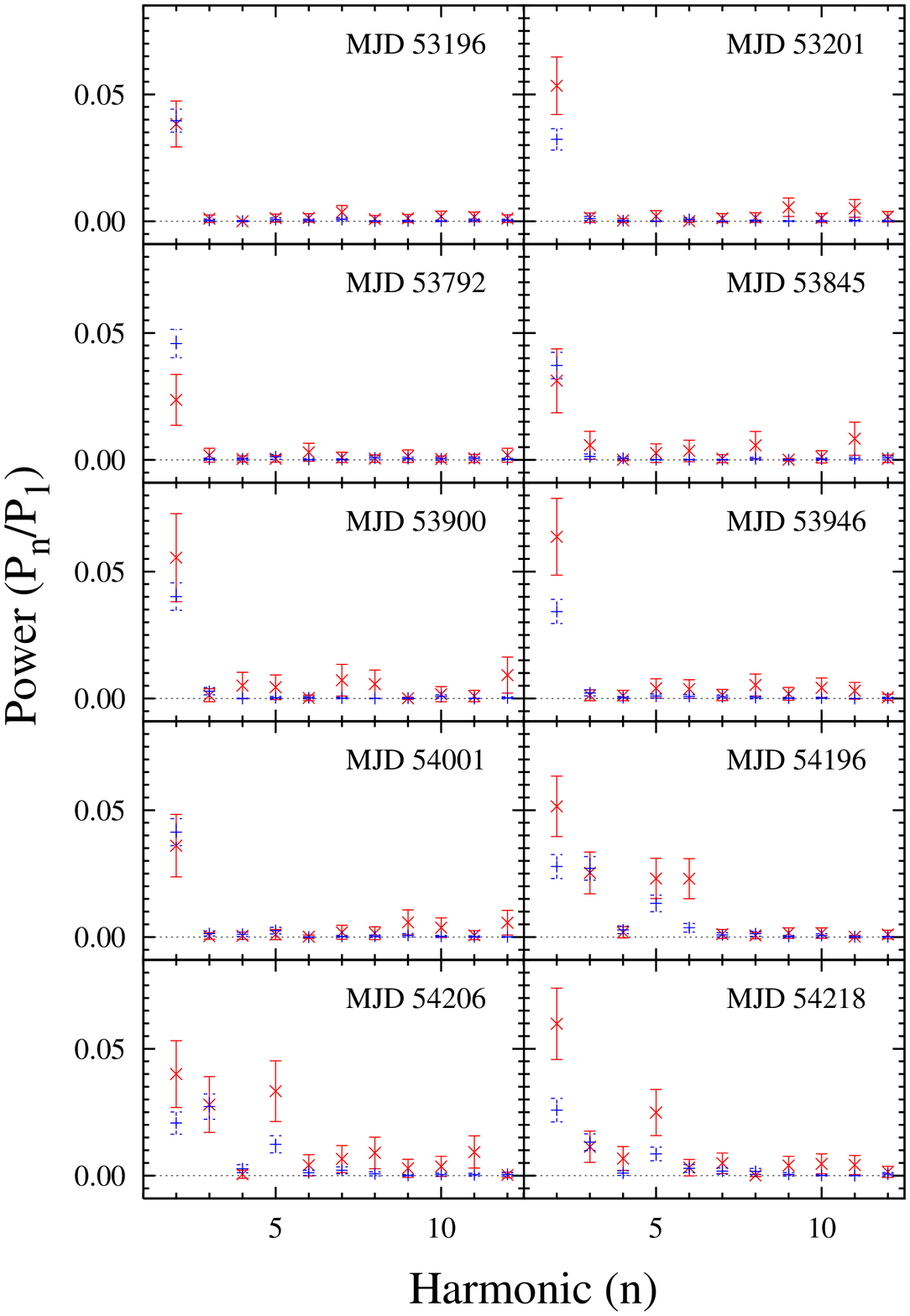} 
\figcaption{\textit{Left}: 1$-$3~keV (\textit{blue}) and 3$-$10~keV
  (\textit{red}) normalized pulsed profiles as observed by \cxo. The
  profiles are phase aligned, have had their minimum bins subtracted,
  and are normalized such that the area underneath them is
  unity.  Note that the 2007 event occured around MJD~54183.  
  \textit{Right}: 1$-$3~keV (\textit{blue}) and 3$-$10~keV
  (\textit{red}) Fourier decomposition of the pulse profiles. The
  powers $P_n$ are plotted in terms of their ratio to the fundamental
  $P_1$. The $n=0$ and the $n=1$ (fundamental) components are
  excluded. Notice how the pulse profile has additional structure
  after the glitch/flux enhancement. \label{fig:profiles}}
%\end{figure} %%% MANUSCRIPT
\end{figure*}

For the \cxo\ observations, we barycentered our corrected Level 2
event lists and then extracted background-subtracted light curves in
the different bands using the same source and background regions as
for the spectral analysis. We folded these lightcurves at the optimal
frequency as determined by a periodogram. The frequencies we obtained
all agreed with the ephemerides determined with our contemporaneous
\xte\ data. Figure \ref{fig:profiles} (\textit{left}) displays
the \cxo\ pulse profiles in the 1$-$3 and 3$-$10~keV bands. Notice
that after the flare (last 3 panels in Figure~\ref{fig:profiles}
\textit{left}), the pulse profile changed from single to at least
triple peaked.  To study the pulse morphology evolution quantitatively
we decomposed the pulse profiles into their Fourier components. In
Figure~\ref{fig:profiles} (\textit{right}) we plot the Fourier
components in terms of their ratio to the fundamental (the $n=0$ and
fundamental are excluded from the plot).  Notice that before the flare
the profile consisted only of the fundamental and an $n=2$ Fourier
component. After the flare, the profile exhibited significant Fourier
components of order $n=2,3,5$.  Interestingly, despite the drastic
changes in the other Fourier components, the $n=4$ component remains
consistent with zero throughout the flare.

%\begin{figure} %%% MANUSCRIPT
 %%% EMULATEAPJ

Similarly, for the \swift\ observations we used the barycentered,
cleaned, and pointed PC-mode data events.  The XRT WT data consisted
of too few pulses to be useful. \swift\ PC-mode data have only
2.5-s resolution, providing fewer than 3 pulse phase bins for this
6.45-s pulsar.  We attempted to improve on the coarse sampling by
utilizing advanced ``bin splitting'' techniques (G. Israel, private
communication), however we find only marginal evidence for a
pulse profile change from these data \citep[but see][]{ci07}.

Using our folded pulse profiles we measured root-mean-square (rms)
pulsed fractions\footnote{For a simple signal of the form $c(\phi) =
A\sin(2\pi \phi) + B$, the rms pulsed fraction is given by
$PF_{\mathrm{rms}}=\frac{1}{\sqrt{2}}\frac{A}{B}$, and the
peak-to-peak pulsed fraction is given by
$PF_{\mathrm{pp}}=\frac{A}{B}$. We opted to use the rms with Fourier
filtering (Archibald et al., in preparation) because it is a more
robust estimator of the
pulsed fraction than peak-to-peak.} using the method described in
\citet{wkt+04}. Our 2$-$10~keV rms pulsed fractions are listed in
Table~\ref{tab:obs} and plotted in Figures~\ref{fig:timing} and
\ref{fig:xrayspec}.  The \xmm\ pulsed fractions, taken from
\citet{gkw06}, were extracted in the same fashion.  Because of
the low time resolution of the \swift\ data, the rms pulsed fraction
would be artificially reduced if the pulse contained as much
harmonic content as the \cxo\ data during the flare. Thus, for each
\swift\ observation we simulated, given its observed pulsed fraction
and mean count rate, what its true pulsed fraction would be assuming
a pulse profile with as much harmonic content as the nearest \cxo\
observation. On average we found a reduction of no more than
$\sim$73\%.  Note in Figure~\ref{fig:xrayspec} the observed tight
correlation between total flux and pulsed fraction \citep[see
also][]{tmt+05} which we discuss in detail in \S~\ref{sec:pfvsf 
disc}. 

\subsection{\hst}
\label{sec:hst}

\begin{figure}
%\epsscale{0.75} %%% MANUSCRIPT
\plotone{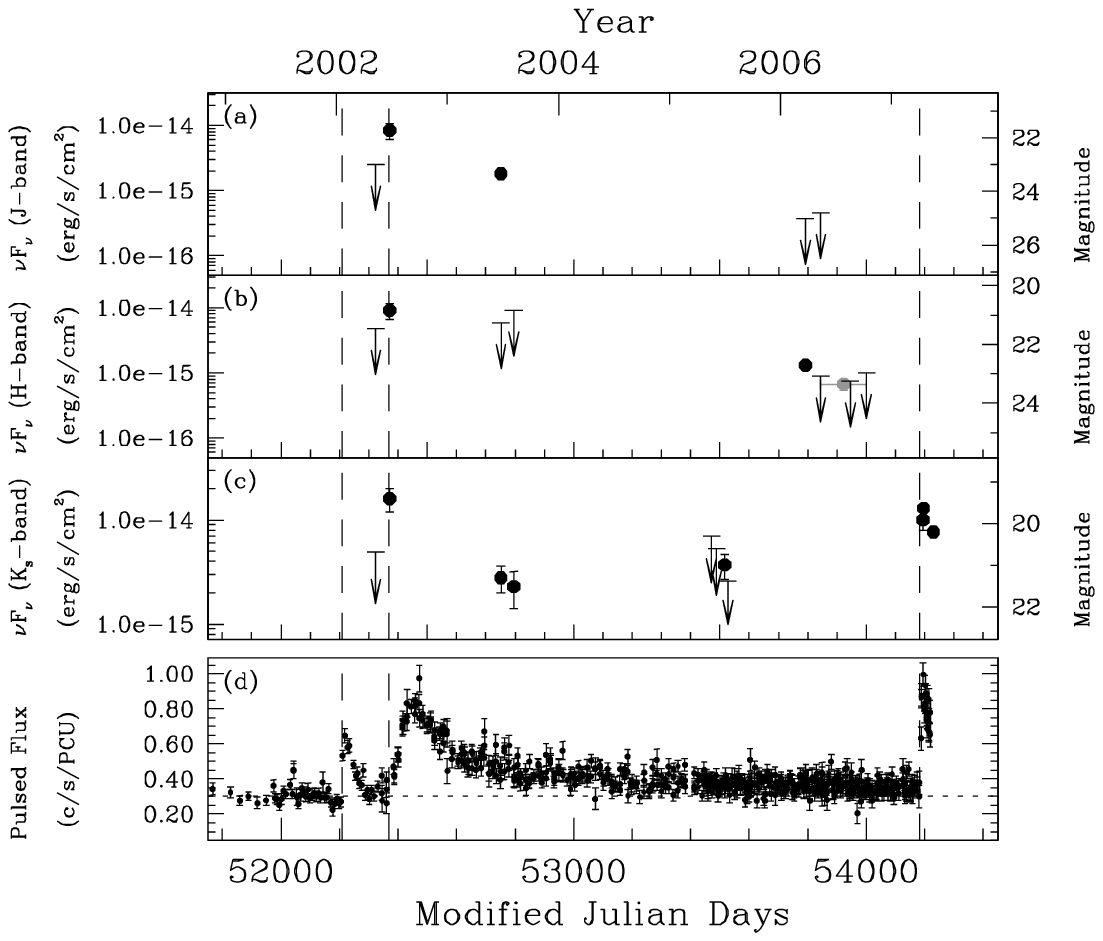}
\figcaption{The evolution of \psr's near-IR brightness. \vlt\ and
  \hst\ results from this work fall after 2004, while data prior to
  that are from previous literature \citep{wc02,ics+02,dv05a}. We also
  show the corresponding magnitude on the right axis; however the
  translation is not exact due to \hst's unique filters.
  Vertical dashed lines indicate the beginnings of the flux flares
  (see Fig.~\ref{fig:timing} caption),
  while horizontal dotted lines indicate quiescent values.  
  (\textit{a}) ``$J$-band'' flux $\nu F_{\nu}$.
  (\textit{b}) ``$H$-band'' flux $\nu F_{\nu}$.  The detection made
  from combining three \hst\ observations (see \S~\ref{sec:hst}) is
  shown in grey. 
  (\textit{c}) ``$K_S$-band'' flux $\nu F_{\nu}$. Results from 2007
  are described in detail elsewhere \citep{wbk+07}.
  (\textit{d}) For reference, we show the 2$-$10~keV rms pulsed flux
  as observed by \xte. 
  \label{fig:nearir}}
\end{figure}

We observed \psr\ using \textit{HST} simultaneously with \textit{CXO}
on four occasions in 2006: February 26, April 20, July 30, and
September 23 (Program ID 10761). Observations were made with the Near
Infrared Camera and
Multi-Object Spectrometer (NICMOS) instrument, a $256 \times 256$
square pixel HgCdTe detector.  Combined with camera 3, it provided a
focus ratio of $f$/17, a field of view of $51' \times 51'$, and a
plate scale of $0\farcs2$~pixel$^{-1}$.  The detector was read out in
MULTIACCUM mode, and a spiral dither pattern with 5$''$ spacings was
applied.  We observed using the filter F110W (similar to the
ground based $J$-band filter) on two occasions and F160W (similar to
$H$-band) on four occasions.  The average FWHM of the point spread
function (PSF) was $\sim0\farcs4$ for all NICMOS observations.
Observing parameters are summarized in Table~\ref{tab:irobs}.

The data underwent ``On-The-Fly-Reprocessing'' involving standard
pipeline routines before being retrieved from the \textit{HST}
Archive.  The calibration routine \texttt{calnica} performed basic
data reduction steps, and \texttt{calnicb} produced a single final
image for each observation, combining the dithered mosaic input
images.  DAOPHOT \citep{ste87} for IRAF v2.12.2 was used to perform
PSF photometry.  Absolute calibration of our measurements was done
using the NICMOS Photometric
Keywords\footnote{\texttt{http://www.stsci.edu/hst/nicmos/performance/photometry}}
for which the uncertainties are believed to be less than 5\%; we found
the aperture correction to correspond with the calibration keywords
by simulating PSFs with the Tiny
Tim\footnote{\texttt{http://www.stsci.edu/software/tinytim/tinytim.html}}
software package.

In order to determine limiting instrumental magnitudes in each of the
observations, we performed PSF photometry on a set of a hundred trial
images to which a single artificial star was added. This artificial
star was placed in a $0\farcs3$ box in a blank region near the
nominal position of the IR counterpart. By varying the
brightness of the star, we define the $3\sigma$ detection limit at the
instrumental magnitude for which the PSF photometry recovers the
artificial star with an uncertainy of 0.3\,mag \citep[see][]{hvvk00}.

At the position of \psr, as measured by \citet{wc02} in $K_S$-band
with the \textit{Magellan} telescope, we find one point source in the
F160W image from 2006 February 26 only.  To confirm the positional
coincidence of this source with the \textit{Magellan} object, we
astrometrically tied $N=31$ nearby field stars in our \hst\ image to
those from the \textit{Magellan} image.  A $\sim0\farcs05$ offset was
found for the \psr\ candidate.  The \textit{Magellan} source's
relative positional uncertainty, given its $0\farcs4$ FWHM radius, is
roughly FWHM/$\sqrt{N}\sim0\farcs07$. Therefore, we consider this
\hst\ object the counterpart of \psr.

Thus, we measure our lone detection of \psr\ to have
\hsth~$=22.70\pm0.14$~mag; uncertainties indicate the nominal errors
determined by DAOPHOT and incorporate PSF fitting uncertainties.
Limiting magnitudes for the three latter \hst\ F160W observations in
which no counterpart was seen are presented in Table~\ref{tab:irobs}.
To deepen the sensitivity, geometric transforms were computed to
register and combine the three non-detection datasets.  In the
combined data, we tentatively detect an extremely faint counterpart of
brightness \hsth\ $= 23.42 \pm 0.20$~mag, where the 3$\sigma$ upper
limit is $>23.5$~mag.  We find no F110W counterpart down to a
3$\sigma$ upper limit of \hstj\ $>25.0$~mag and $>24.8$~mag, on
February 26 and April 20, respectively.  Results are plotted in
Figure~\ref{fig:nearir}. 

To examine how this compares with past results, we also calibrated
our measurements using the $H$ magnitudes from \citet{dv05a}.  A
relative offset is found between \hst's F160W filter and ground based
$H$-band, as a function of $J-H$ color.  We know the color of \psr\
only during a brightened state \citep[$J-H\approx 0.9$ in April
2002;][]{wc02}, and suspect that it is not constant; therefore, this
is not a perfect comparison, but provides a rough estimate.  Based on
the magnitudes of seven nearby field objects, we measure $H\approx
22.73$~mag on February 26, which is very near what we find for
\hsth.

%\begin{deluxetable}{lcccccc} %%% MANUSCRIPT
\begin{deluxetable*}{lcccccc} %%% EMULATEAPJ
\tablecaption{\hst\ and \vlt\ observing parameters and results
  \label{tab:irobs}}
\tablehead{
\colhead{Date} & \colhead{MJD} & \colhead{Exposure} &
\colhead{Filter\tablenotemark{a}} & \colhead{Limiting} &
\colhead{Detection} & \colhead{$\nu F_{\nu}$\tablenotemark{b}}
\\
\colhead{} & \colhead{} & \colhead{(s)} & \colhead{} &
\colhead{Magnitude} & \colhead{} & \colhead{(10$^{-15}$ ergs s$^{-1}$ cm$^{-2}$)}
}
\startdata
\multicolumn{7}{c}{\vlt\ Observations}\\
\hline
2005 Apr 11 & 52324 & 900 & $K_S$ & $>$20.3 & \nodata & $<$7.0\\
2005 Apr 29 & 53471 & 900 & $K_S$ & $>$20.6 & \nodata & $<$5.3\\
2005 May 27 & 53489 & 900 & $K_S$ & $>$21.2  & 21.0(3) & $<$3.1; 3.7$\pm$1.0\\
2005 Jun 6  & 53527 & 900 & $K_S$ & $>$21.4 & \nodata & $<$2.6\\
\hline
\multicolumn{7}{c}{\hst\ Observations}\\
\hline
2006 Feb 26 & 53792 & 2637 & F110W & $>$25.0 & \nodata   & $<$0.37\\
2006 Feb 26 & 53792 & 1196 & F160W & $>$23.7 & 22.70(14) & $<$0.49; 1.3$\pm$0.2\\
2006 Apr 20 & 53845 & 2637 & F110W & $>$24.8 & \nodata   & $<$0.45\\
2006 Apr 20 & 53845 & 1196 & F160W & $>$23.1 & \nodata   & $<$0.89\\
2006 Jul 30 & 53946 & 1037 & F160W & $>$23.3 & \nodata   & $<$0.74\\
2006 Sep 23 & 54001 & 1037 & F160W & $>$22.9 & \nodata   & $<$1.0 
\enddata

\tablecomments{Magnitude uncertainties reflect errors determined by
  DAOPHOT; upper limits are 3$\sigma$ limits, as defined in the text.
  Measured quantities are ``observed'', ie. have not been corrected
  for reddening/extinction effects.}
\tablenotetext{a}{Although the filter system of \textit{HST} does not
  precisely match standard ground-based near-IR filters, note that
  F110W $\approx J$-band and F160W $\approx H$-band.}
\tablenotetext{b}{To convert standard IR magnitudes to flux, we take
  as the $K_S=0$ mag zero point $\nu F_{\nu} = 9.28 \times 10^{-7}$
  ergs s$^{-1}$ cm$^{-2}$, derived from \citet{cox00}.  \hst\ fluxes
  are determined from NICMOS Photometric Keywords.}
%\end{deluxetable} %%% MANUSCRIPT
\end{deluxetable*} %%% EMULATEAPJ

\subsection{\vlt}
\label{sec:vlt}

We have analyzed archival observations of \psr\ which were obtained
with NAOS-CONICA (NACO), the near-IR adaptive optics (AO) instrument
at \vlt\ of the European Southern Observatory. The source was observed
for 900\,s in $K_\mathrm{s}$-band using the S27 camera
($28\arcsec\times28\arcsec$ field-of-view and 27 mas\,pix$^{-1}$ pixel
scale) on four occasions (April 11, April 29, May 27 and June 6) in
2005. Star A \citep{wc02}, located $9\farcs5$ from the AXP, was used
as the wavefront-sensing (WFS) star. Due to the relative faintness of
the WFS star ($r'=16.5$~mag), the AO correction was only partial.

The observations, either dithered sets of 5 $2\times90$\,s
integrations (on the first night) or of 10 $1\times90$\,s 
integrations, were corrected for dark current, flat-fielded (using
averages derived from the science frames) and corrected for variations
in the sky (using median averages of the flat-fielded science
frames). Finally, the reduced science images were aligned using
integer pixel offsets and median-combined to create an average image
for each night.

Using DAOPHOT, we tried PSF photometry on each of the four averaged
images, using different models for the PSF.  Because the AO correction
is only partial, we found that the PSF is reasonably well modeled by
a Moffat function of exponent 2.5 in combination with a look-up
table. The width of the PSF that was determined for each observation
was $0\farcs34$ (April 11), $0\farcs43$ (April 29), $0\farcs24$ (May
27) and $0\farcs20$ (June 6). Only in the observation of May 27 did we
detect the near-IR counterpart to \psr.

We calibrated the instrumental magnitudes and magnitude limits using
stars X2 to X8 and A and B against the calibrated $K_\mathrm{s}$
magnitudes of Wang \& Chakrabarty (2002). The rms uncertainty in the
calibration is about 0.2\,mag. The near-IR counterpart to
1E\,1048.1$-$5937 was detected at $K_\mathrm{s}=21.0\pm0.3$ on 2005
May 27; the $3\sigma$ upper limits for April 11, April 29, May 27 and
June 6 are $K_\mathrm{s}>20.3$, $K_\mathrm{s}>20.6$,
$K_\mathrm{s}>21.2$ and $K_\mathrm{s}>21.4$, respectively.  Limiting
magnitudes were determined with the same method described in
\S~\ref{sec:hst}.  Results are shown in Figure~\ref{fig:nearir}.

Note that post-2007 event optical and near-IR observations of \psr\
were obtained with \vlt\ and \textit{Magellan}
\citep{wbk+07,ict+07}. Preliminary results of the $K_S$-band
observations are shown in Figure~\ref{fig:nearir}, and
details on some of the observations can be found in \citet{wbk+07}.

\section{Discussion}
\label{sec:discussion}

Explaining the origin of the different types of AXP variability across
the electromagnetic spectrum is a major challenge for the magnetar
model.  Many fundamental questions remain open.  In the case of \psr,
how are timing instabilities related to radiative variability?  AXP
spin-up glitches sometimes are accompanied by major radiative events,
as in the 2002 outburst of AXP \tfn\ \citep{kgw+03}, and
sometimes not, as in the large glitch of 1E~1841$-$045 \citep{dkg07}.
Moreover, are the \textit{gradual} large-scale changes seen in the
2001-2002 flares of \psr\ of a similar origin to AXP outbursts like
that of \tfn\ or XTE~J1810$-$197 \citep{ims+04}, and how can they
be accounted for by magnetar models?  Also, what is the nature
of the optical/IR emission observed in five out of eight confirmed
AXPs?  Suggested origins include coherent or curvature
emission in the plasma magnetosphere \citep{egl02,bt07}, but so far
this has not be confirmed.  Are the low- and high-energy emission
mechanisms intimately connected \citep{hh05b}?  Correlated post-burst
flux decay in the X-ray and near-IR regimes has been seen in at least
one other AXP \citep{tkvd04}.

Given the many observational properties that have been characterized for
\psr\ in this work and for other AXPs in similar studies, which property
will emerge as being the most constraining of physics is hard to know.
The most promising behaviors are those which show clear correlations
with others, or those that are common to many or all AXPs and SGRs.
In this section, we discuss the behavior of \psr, and focus on the
phenomena that are potentially the most useful for testing the magnetar
or other competing models.  We first consider the quiescent phase we
have observed in 2004-2006, and its implications for AXP models, then
subsequently discuss the source's return to activity in 2007 March.

\subsection{The 2004-2006 Quiescent Phase}
\label{sec:quiescent discussion}

\subsubsection{X-ray Flux and Spectrum}
The X-ray flaring observed pre-2004 contrasts strongly with the stable
pulsed fluxes we observed in 2004-2006
(Fig.~\ref{fig:timing}\textit{c}).  Clearly the source's
stable state 
is also its faintest.  The X-ray spectrum during this quiescent phase
is also fairly constant, with $kT$ evolving post-flare, on a time
scale of several years.  Interestingly, $kT$ does not return to its
pre-flare value as measured in 2000 December (see
Fig.~\ref{fig:xrayspec}\textit{a}). Similarly, the photon index
(Fig.~\ref{fig:xrayspec}\textit{b}) 
decreased in 2004-2006, away from its softer pre-flare value.
Meanwhile the pulsed fraction (Figs.~\ref{fig:timing}\textit{e} and
\ref{fig:xrayspec}\textit{e}) was slowly rising in 2004-2006, as 
the source grew fainter, as if, unlike $kT$ and $\Gamma$, it was
slowly recovering to the pre-flare value.  Thus, overall, the
quiescent period is characterized by slow evolution, on timescales of
years, in which the source flux and pulsed fraction slowly relaxed
back to their pre-flare values, while the source spectrum varied
significantly (Fig.~\ref{fig:xrayspec}\textit{c}) and did not relax
back to its pre-flare state.  This is suggestive of a fixed energy
loss rate in quiescence, though perhaps with a different
magnetospheric current configuration, which impacts the surface
emission via return currents \citep{tlk02}. We note, on the other
hand, that the 2006 and pre-2001 pulse profile were very similar if
not identical \citep{kgc+01}, and that the \citet{tlk02} simple
prediction for the decay time of the sort of magnetospheric twist
required for \psr\ is an order of magnitude too large for standard
parameters. The evolution could also be purely due to the thermal
component \citep{og07}.

\subsubsection{Timing Stability}
Simultaneous with the slow evolution of the X-ray flux, pulsed
fraction, and spectrum, the pulsar's rotational behavior clearly
stabilized, with the source spinning down relatively smoothly at a
value close to the long-term average in 2004-2006
(Fig.~\ref{fig:timing}\textit{a}). Whereas previously this AXP
distinguished itself from others by defying attempts at phase-coherent
timing and exhibiting large torque variations
(Fig.~\ref{fig:timing}\textit{b}), in 2004-2006, \psr\ resembled,
from a timing point of view, other AXPs which are relatively stable
rotators, at least when not glitching
\citep[e.g.][]{kcs99,dkg07}. This behavior demonstrates a clear
relationship between timing and radiative properties, and that the
``noise'' seen in the frequency derivative during 2001-2004 is
likely physically different from the ``timing noise'' observed
ubiquitously in radio pulsars and in otherwise radiatively stable
AXPs.

In the context of the magnetar model, magnetospheric activity can
account for both torque and X-ray flux variability, although the
former is most sensitive to currents anchored closest to the magnetic
poles, so that only a broad correlation between flux and torque is
expected \citep{tlk02} and indeed is seen when considering many AXPs
and SGRs \citep{mw01}.  That the torque and luminosity in \psr\ do not
vary simultaneously or in a clearly correlated way is thus not
necessarily problematic if the magnetospheric current configuration is
changing, although it does demonstrate that the magnetar model in this
particular regard is not strongly predictive.

Although the magnetar model is favored, it has also been suggested
that accretion from a fossil debris disk could explain AXP spin
characteristics \citep{chn00,alp01} as well as all aspects of the
broad band emission \citep{ec06}.  While it is now evident that
accretion alone cannot be responsible for all observed properties,
most notably the energetic X-ray bursts, a ``hybrid'' model has been
invoked that puts a thin debris disk around a highly magnetized pulsar
\citep{ea03}.  In this hybrid case, both the persistent luminosity and
pulsar spin down are related to the mass transfer rate, $\dot{M}$.
According to \citetalias{gk04}, who compare \psr's torque changes with
changes in \xte\ pulsed flux, the scale of their variability does
not obey an expected relationship for an accreting pulsar undergoing
spin-down, thus presenting a challenge to models of fossil disk
accretion.  It could be argued, however, that since only the pulsed
flux $P_X$ was being monitored, the total X-ray luminosity $L_X$ is
still an unknown quantity.  We address this by using the quantitative
correlation we establish between pulsed fraction
$\mathcal{P}_{\mathcal{F}}$ and total unabsorbed flux $F_X$ (discussed
in \S~\ref{sec:pfvsf disc} and shown in Fig.~\ref{fig:pulsefrac}) to
simulate a well sampled set of phase-averaged flux data as a function
of \xte\ pulsed flux $P_X$, given that $\mathcal{P}_{\mathcal{F}}$
is simply $P_X/F_X$.  The resulting ``new'' 
total flux, $F_X$, is shown in Figure~\ref{fig:timing}\textit{d}, along
with $\dot{\nu}$ in Figure~\ref{fig:timing}\textit{b} for comparison.
At its most variable in 2002-2004, the absolute value of $\dot{\nu}$
changed by a factor of $>$10 in less than a year, while the maximal
change in total unabsorbed flux is a factor of $\sim$6 from peak
to quiescence.  Note that these variations are not simultaneous.
For a pulsar experiencing a spin-down torque due to mass accretion
while in a quasi-equilibrium ``tracking'' phase (i.e. the AXP phase),
a strong correlation described by $L_X \propto |\dot{\nu}|^{7/3}$ can
be derived from \citet[][equation 3]{chn00}, where the magnitudes of
torque and $\dot{\nu}$ are proportional, the radius of the magnetosphere
is defined by the Alf\`en radius which is also a function of $\dot{M}$,
and $L_X \propto \dot{M}$.  Thus, a factor of $>$10 change in $\dot{\nu}$
should be reflected by a factor of $>$200 simultaneous change in $L_X$,
and thus $F_X$, if this model is correct.  Clearly this is not observed,
rendering this particular accretion scenario unlikely.  This point is
further emphasized by the significant time offset between the changes in
torque and flux.  Overall, the fact that \psr's long history of highly
irregular spin-down is mirrored by trends in its X-ray emission only
in a broad, rather than strict, sense strongly suggests that active
accretion is not happening in this case.

\subsubsection{Near-IR Quiescence}
Our original purpose for making simultaneous \cxo/\hst\ observations
was to compare low-level near-IR and X-ray flux and spectral changes,
with each observed with the same instrument, in order to look for
correlations.  Given that the near-IR source had faded considerably
in the \hst\ observations, rendering it only marginally detected in
2006, this was not possible.  Compared to the handful of near-IR
detections made throughout 2002-2003 (see Fig.~\ref{fig:nearir}) it
is clear that 2005-2006 marked a period of near-IR quiescence in
\psr. At the time of the 2005 \vlt\ observations in which the AXP was
faintly seen in $K_S$-band, its flux was consistent with the last
detection in mid-2003.  With \hst\ in 2006 we find that its $H$ and
$J$ magnitudes have dropped lower than ever before observed.
There is marginal evidence for near-IR flux variability during
quiescence on comparably short timescales to the variability in
X-ray spectral parameters, although closely spaced near-IR monitoring
observations are required to confirm this. 

That the near-IR faded roughly in concert with the X-ray flux is
notable, suggesting a correlation similar to that seen in AXP
\tfn\ \citep{tkvd04}.  This is
discussed further below.  However such a correlation can be argued to
be expected in both the magnetar and disk models.  In the magnetar
model it would imply that the near-IR emission is magnetospheric and
hence varies, as do the X-rays, when the magnetospheric configuration
varies \citep{hh05b,bt07}. In the disk model, a correlation between
near-IR emission and X-ray flux is naturally expected since the
putative disk is heated via X-ray illumination \citep{ega06,wck06},
however why the torque should have varied strongly and
non-simultaneously is a puzzle (see above). 

In any case, the concurrence of the near-IR and X-ray quiescence and
the timing stability is unlikely to be an accident.  Similarly, the
presence of the two flares and the subsequent strong torque changes
are also not likely to be by chance.

\subsection{The 2007 March Event and its Aftermath}
\label{sec:flare discussion}

\psr\ reactivated in 2007 March, when a sudden pulsed flux increase
was observed (Fig.~\ref{fig:timing}\textit{c}), accompanied by a
large spin-up glitch, the details of which will appear elsewhere (Dib
et al., in preparation). Our follow-up X-ray observations with \cxo\
and \swift\ show that the spectral parameters, pulse shape, total
flux, and pulsed fraction are drastically changed since this flare
occurred (Figs.~\ref{fig:xrayspec} and \ref{fig:profiles}).  In
particular, the total X-ray flux 
was $>$7 times greater in the 2$-$10~keV band compared with
quiescence, while the pulsed fraction decreased from $\sim$75\% to
$\sim$20\%.  Compared to the previous long-term flares from \psr, the
onset of this event took place more quickly by a factor of $>$4, the
peak pulsed flux is $\sim$10\% greater, and the ``high'' state appears
to be lasting longer.  Optical and near-IR follow-up
with the \textit{Magellan Telescope} and \vlt\
\citep{wbk+07,ict+07} reveal an increase in both $I$ and $K_S$-band
of $\sim$1.3 mag.  Very recently, \citet{rtic07} reported on \xmm\ ToO
observations, and revealed that nearly three months after the flare
onset, the total X-ray flux has decreased slightly but is still
$\sim$5 times brighter than in quiescence.

\subsubsection{Near-IR Enhancement}
Previously, an anti-correlation between near-IR and X-ray flux in this
AXP was suggested, based on the correspondence of the highest near-IR
detection with a low point in X-ray pulsed flux\footnote{Note that an
error in Figure~4 of \citet{dv05a} places the brightest $K_S$ flux on
the wrong date; we have corrected this in our
Figure~\ref{fig:nearir}\textit{c}.} \citep{dv05a}. Given the new
measurements from both before and after the recent event, such an
anti-correlation is highly questionable.  The significant rise in
near-IR flux now appears correlated with the most recent X-ray
flare. Interestingly, in 2002, a high near-IR detection was {\it
between} the two X-ray flares; this suggests that the near-IR emission
is not exactly correlated with the X-rays, but rather an enhancement
can precede, follow, or last longer than an X-ray enhancement.  A
near-IR brightening preceding an X-ray enhancement would rule out
illuminated disk models for this source; frequent mid- or near-IR
observations are required to check this.  As discussed above, AXP
\tfn\ also demonstrated strongly correlated near-IR and 
X-ray flux decay following its large 2002 outburst \citep{tkvd04}, but
in contrast, 4U~0142+61 was highly variable in the near-IR despite
X-ray stability \citep{dv06c}, and has shown no evidence of near-IR
changes coinciding with X-ray bursts (Gonzalez et al., in
preparation).  Note that the claim of correlated flux decay in the
case of XTE~J1810$-$197 \citep{rti+04} is under dispute
\citep{crp+07}.  Such inconsistent behaviour is puzzling for both the
magnetar and disk models.

\subsubsection{Correlation between Pulsed Fraction and Total Flux}
\label{sec:pfvsf disc}

%\begin{figure} %%% MANUSCRIPT
\begin{figure*} %%% EMULATEAPJ
%\epsscale{1.0} %%% MANUSCRIPT
\plottwo{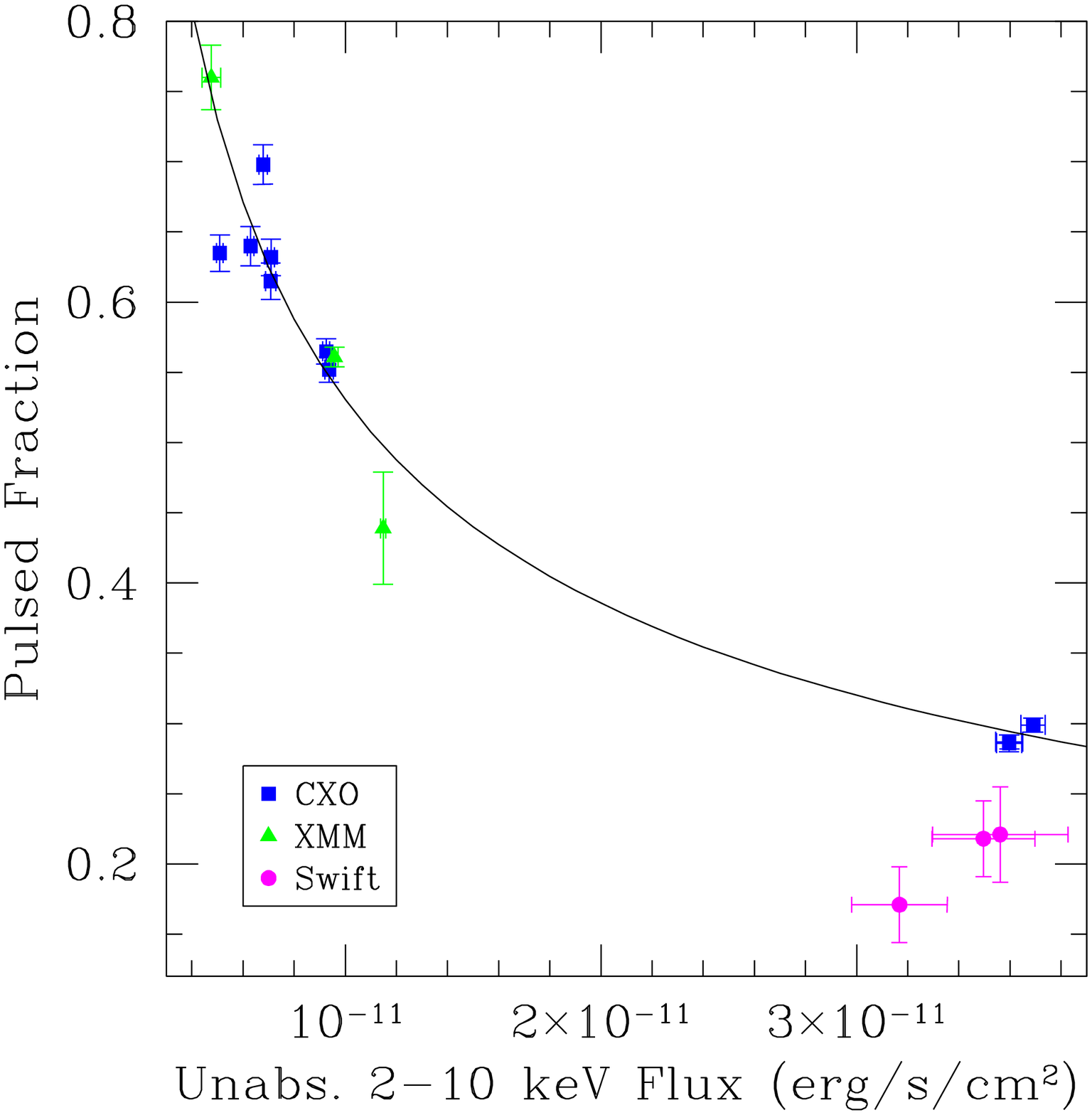}{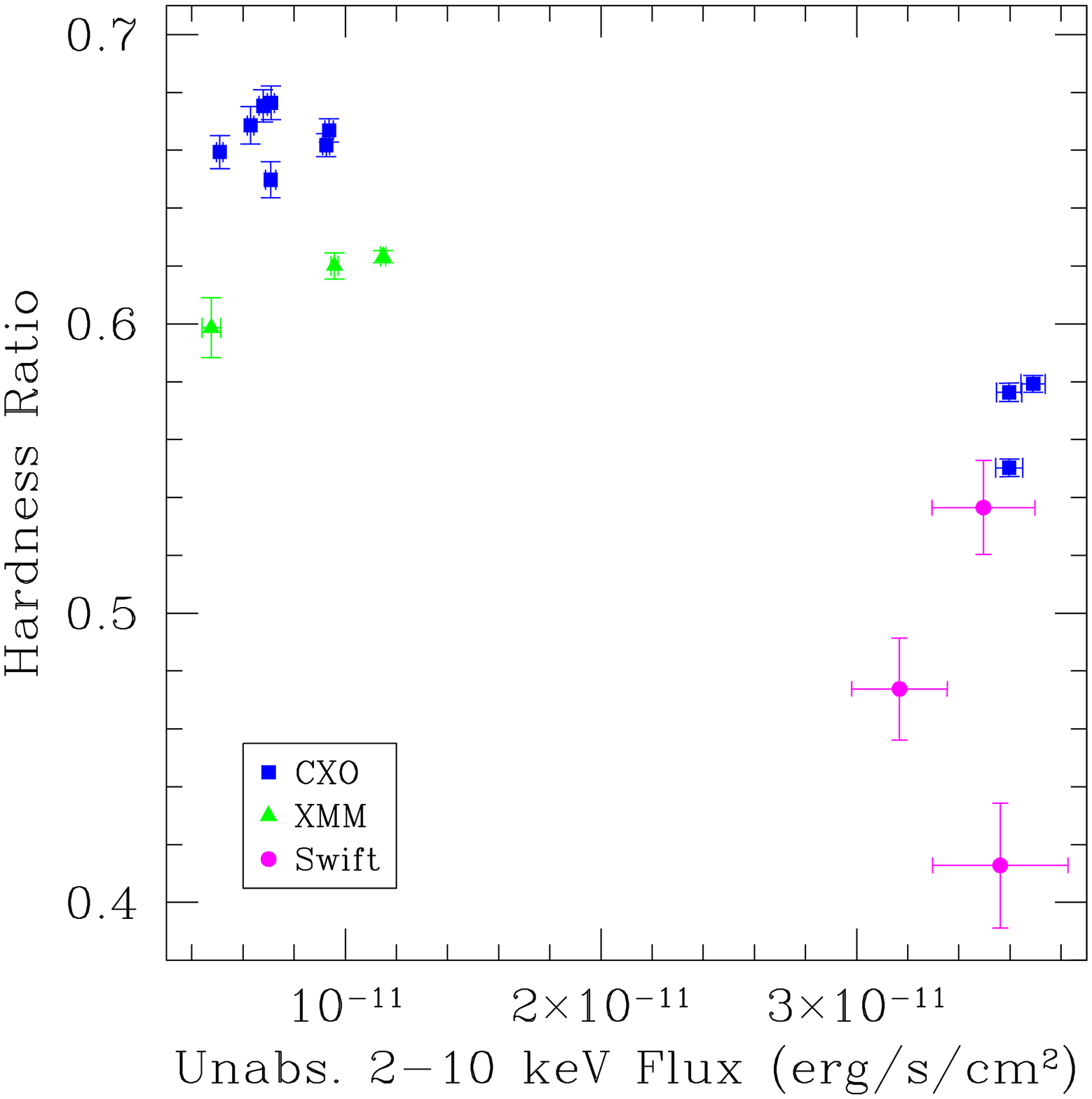}
\figcaption{X-ray flux dependent properties of \psr, from 6 years of
  observations with \cxo\ (\textit{squares}), \xmm\
  (\textit{triangles}), and \swift\ (\textit{circles}). \textit{Left:}
  Pulsed fraction vs. total unabsorbed flux, both in 2$-$10~keV.  The
  solid black line shows the best-fit power law that describes the
  correlation between the two quantities. \textit{Right:} Hardness
  ratio vs. total unabsorbed 2$-$10~keV flux.  The hardness
  ratio is defined as $(S-H)/(S+H)$ where $S$ and $H$ are the
  phase-averaged count rates in the 1$-$3 and 3$-$10~keV bands,
  respectively. \label{fig:pulsefrac}}
%\end{figure} %%% MANUSCRIPT
\end{figure*} %%% EMULATEAPJ

One conclusion to be drawn from
Figures~\ref{fig:timing}\textit{d} and \ref{fig:timing}\textit{e} is
that \psr's pulsed fraction $\mathcal{P}_{\mathcal{F}}$ is strongly
anti-correlated with the total X-ray flux $F_X$, a trend already
noticed by \citet{tmt+05} and \citet{gkw06}.  Now, with our larger
sample of data covering a wider dynamic range, we can quantify this
correlation: we find a power-law dependence of
$\mathcal{P}_{\mathcal{F}} \propto F_X^n$ where $n = -0.46 \pm 0.02$
(see Fig.~\ref{fig:pulsefrac}), with $F_X$ in the 2$-$10~keV range. In
fitting, we set the index as a free parameter, ie. not tied
to any model; therefore, some scatter is expected, as evident in the
large $\chi^2_{\nu} = 9.36$ produced by the fit ($\nu = 14$
dof). However, in finding the uncertainty on $n$, we have scaled the
$\mathcal{P}_{\mathcal{F}}$ data uncertainties to force $\chi^2_{\nu}
= 1$, effectively scaling up $\sigma_n$ as well. A similar correlation
has been proposed in another AXP, 1RXS J170849.0$-$400910
\citep{dkg07}, although in that case it is such that the pulsed flux 
remains nearly constant in the presence of $>$50\% total flux
changes \citep{cri+07}.

How could an increase in phase-averaged flux be met with a
simultaneous decrease in pulsed fraction?  A growing hot spot on the
neutron-star surface, a result of either a changing magnitude or
configuration of returning magnetospheric currents or internal
processes, could at least in principle produce such an effect, if the
initial hot spot size were large enough that its size as viewed from a
distant observer were a significant fraction of the stellar surface.
We note that \citet{og07} suggest that magnetar afterglows may be
dominated by surface thermal, rather than magnetospheric, changes;
detailed simulations of AXP pulsations using gravitational light
bending and appropriate radiation beaming functions
\citep[e.g.][]{dpn01} in addition to magnetospheric scattering
\citep{tlk02,ft07} are needed to see if the observed correlation can
be reproduced.

\subsubsection{Correlation between Hardness and Total Flux}
A prediction made by the twisted magnetosphere model of \citet{tlk02}
is that enhanced emission should also be spectrally harder, since both
are associated with larger magnetospheric twist angles.  Indeed, we
observed a correlation between the total X-ray flux and spectral
hardness: see Figure~\ref{fig:pulsefrac}.  Similar behaviour
has also been reported in 1RXS~J170849.0$-$400910
\citep{roz+05,cri+07}.  This can be interpreted as a confirmation of
an important magnetar model prediction, namely that both the spectral
hardness and the total flux should increase for an increasing
magnetospheric twist angle \citep{tlk02}.  Hardening of the spectrum
might also be an expected effect of increased luminosity if a large
injection of thermal seed photons is repeatedly 
up-scattered due to resonant cyclotron scattering feedback processes
in the magnetosphere, thereby shifting the photon energies higher.
Recently \citet{og07} have suggested that the hardness-intensity
correlation in magnetars is a result of purely surface thermal
changes, with no change in the magnetospheric configuration.  It would
be interesting to apply their model to the \psr\ data as they span a
much greater dynamic range in flux than did the sources studied by
\citet{og07}.  This is beyond the scope of our paper, however.

\subsubsection{2007 Pulse Profile Changes}
As seen in the \cxo\ data obtained after 2007 March, the pulse profile
of \psr\ changed abruptly after the X-ray flux enhancement and glitch.
Several new harmonics are clearly visible, and are present in a
largely energy-independent way (Fig.~\ref{fig:profiles}).  This
extra power in the higher harmonics clearly signals a complication of
the surface and/or magnetospheric configuration. 
This change contrasts with that seen after the 2002 outburst
of \tfn, which involved mainly an exchange of powers between the
fundamental and first harmonic, with higher harmonics remaining
relatively unchanged \citep{wkt+04}.  Also, the change contrasts with
that seen during the 1998 SGR~1900+14 giant flare, in which the SGR
pulse profile simplified greatly immediately post-flare
\citep{gkw+02}. However, the opposite occured followed the 2004 giant
flare from SGR~1806$-$20 \citep{pbg+05}, when a sinusoidal-to-complex
evolution in pulse morphology took place, similar to that of \psr.
This diversity of behaviors in magnetars post-outburst is problematic
for understanding the underlying physics; indeed it indicates a wide
variety of phenomenon phase space, which itself must be explained by
models. The peculiar suppression of the 4th harmonic in the
post-glitch \psr\ profile is particularly puzzling to us; perhaps it
indicates an important and unchangeable symmetry in the combined
emission and viewing geometries, or perhaps it is a chance occurence
and will not occur again in future events.

\subsubsection{2007 April Spectral Feature?}
In the \cxo\ observation made on 2007 April 6, the one closest to and
just after the 2007 March event, the X-ray spectrum of \psr\ could not
be well fit by a simple continuum model, in contrast with the other
datasets.  We identified a possible absorption line near 2.7~keV,
though we note that the spectral model including this line is still
not a good fit to the data.  This suggests that other low-level
features may be present as well, including a possible feature at
roughly half the above line's energy, $\sim$1.3~keV.  The putative
2.7~keV line joins the growing list of puzzling and sometimes
statistically marginal AXP/SGR spectral features
\citep{iss+02,isp03,ris+03,roz+05}.  Examples of some feature
detections that are clearly statistically significant are,
interestingly, repeatedly near $\sim$13~keV, e.g.  in \psr\
\citep{gkw02,gkw06}, in XTE~J1810$-$197 \citep{wkg+05}, and very
recently AXP 4U~0142+61 \citep[][and Gavriil et al., in
preparation]{gdkw07}.  The feature we see in the 
2007 April data for \psr\ are unlike any of these.  It is perhaps more 
reminiscent of the absorption features seen in some ``isolated neutron
stars'' \citep[e.g.][]{bcdm03,vkd+04}, and, if similarly interpreted as
an electron cyclotron line, implies a magnetic field of $\sim 3 \times
10^{11}$~G, or $\sim 6 \times 10^{14}$~G for a proton cyclotron line.
The latter value is of course more in line with the field expected
near the surface of a magnetar.  However, the feature could also be an
atomic transition \citep[e.g.][]{mh07}; it is difficult to know given the
present data.  Regardless, the spectrum 10~days later, on 2007 April 16,
showed no such features, indicating that whatever their origin, it had
subsided, and on a shorter timescale than that of  any profile or flux
relaxation. 

\subsubsection{Comparison of the 2007 \psr\ Event with Other Similar
  Events}
The 2001-2002 flares observed in \psr\ are unique in that they had
very well resolved rise times of many weeks \citepalias{gk04}.  The
2007 April event, by contrast, involved an X-ray pulsed flux
enhancement that rose at least 4 times faster.  In this sense it is
perhaps more similar to the 2002 outburst of \tfn\
\citep{kgw+03}, although in that case a brief rise could have been
missed by the sparse monitoring.  During that outburst, \tfn\ 
exhibited an order-of-magnitude increase in pulsed and phase-averaged
flux, as well as numerous other changes including (but not limited
to) $\sim$80 SGR-like bursts.  Similarly to \psr,
both the total and pulsed flux increased during the \tfn\ outburst,
and the pulsed fraction decreased, going from 23\% in quiescence down
to 15\% \citep{wkt+04}. However, the pulsed fraction was not as
tightly correlated to the pulsed flux: the pulsed fraction recovered
within $\sim$3 days \citep{zkw+07}, whereas the flux took several
months to return to 
its pre-outburst value \citep{wkt+04}.  In terms of energetics,
the total energy in 2$-$10~keV released in the \tfn\ outburst was
$3 \times 10^{39}$~erg and $2 \times 10^{40}$~erg during the rapid and
gradual decay components, respectively \citep[see Table~5
of][]{wkt+04}.  For \psr, we estimate that up to 2007 April 28, the
date of the last \cxo\ observation, the total energy emitted is
comparable, roughly $7\times 10^{39}$~erg (2$-$10~keV), based on the
pulsed fraction-total flux correlation and assuming a
distance of 2.7~kpc \citep{gmo+05}, or $8\times 10^{40}$~erg assuming
9.0~kpc \citep{dv06a}.  Near-IR flux variability,
possibly (in the case of \psr) and likely (in \tfn) correlated to
X-ray flux, was observed following both events.  That the 2007 \psr\
event clearly involved a large rotational glitch, as did the 2002
\tfn\ event, is also a commonality, although glitches at the
time of the 2001-2002 flares could easily have been missed because of
the large amount of timing noise and our inability to phase connect
the data at that time (Dib et al., in preparation).  SGR-like bursts
from \psr\ have not been observed thus far in 2007, but that does not
preclude their existence given that our observing duty cycle is so
low. Similar to \psr, \tfn\ exhibited a pulse profile change, albeit
one involving only the fundamental and first harmonic, in constrast to
the appearance of higher harmonic structure in \psr.  Finally, as in
\psr, spectral hardening was observed during the 
\tfn\ flare.  Thus overall, the 2007 \psr\ event has practically all
properties consistent with the 2002 \tfn\ event, but differs from its
earlier flares primarily by rise time.

The transient AXP XTE~J1810$-$197 presents another interesting example
for comparison.  This AXP, which ``turned-on'' in late 2002
or early 2003 \citep{ims+04}, has been steadily declining in X-ray
flux since, reaching apparent quiescence 
in 2005-2006 \citep{gh07}.  Unfortunately, very little is known about
the onset of enhanced emission, ie., whether it was accompanied by a
glitch/burst, or what the timescale of the rise was.  While the
exponential decay timescale of several years does resemble that of
\psr's 2001-2002 post-flare behavior, how the most recent flare will 
fade is as yet undetermined.  Furthermore, the peak and quiescent
fluxes of XTE~J1810$-$197 differed by a much greater amount, nearly 2
orders of magnitude \citep{ghbb04}, while the recent total 2$-$10~keV
flux increase in \psr\ was by a factor of $\sim$7. Spectrally,
XTE~J1810$-$197 was harder in outburst and softer in quiescence,
similar to \psr, when modelled as 2-temperature blackbody \citep{gh07}
or as a blackbody plus power law (W. Zhu, private communication).  One
major distinction, however, is the clear evidence for a 
positive correlation between pulsed fraction and total flux in
XTE~J1810$-$197 \citep{gh07}. While it may be tempting to link the
transience of XTE~J1810$-$197 to \psr-like post-flaring behaviour, the
dissimilarities are important and remind us that we may be comparing
mere coincidences.

Compared to the SGR giant flares, rare events in which an enormous
amount of broad band energy \citep[$10^{44}-10^{46}$~erg;][]{hbs+05} is
output, the \psr\ event is orders of magnitude less energetic.  Giant
flares release most of their energy in a short ($<$1~s), initial
$\gamma$-ray spike, a property not observed in this event, although a
small initial burst could have been missed. Furthermore, the 2007
event is prolonged, and while giant flares do gradually decay, the
amount of energy in the tail is usually small compared to that of the
spike \citep{wkt+04,hbs+05}.

On the other hand, the gradual evolution of the persistent properties
seen before and after the 2004 December giant flare of SGR~1806$-$20
do show some resemblance to \psr's behaviour.  During this pre-flare
period of enhanced burst activity in SGR~1806$-$20 that began in
mid-2003, the pulsar 
torque, pulsed flux, total flux, and hardness increased on a
$\sim$year timescale, peaking several months before the giant flare
occurred in 2004 December, and continuing to decline well after the
flare epoch \citep{wkf+07}.  The gradual changes, particularly the
well-resolved factor of $>$2 increase in total unabsorbed flux,
perhaps resemble the 2001-2002 events of \psr\ more than the 2007
event.  As with \psr, a correlation was found between spectral
hardness and intensity in SGR~1806$-$20, and there appeared to be some
sort of physical connection between frequency derivative (or torque)
evolution and variability in the spectrum and pulsed flux
\citep{mte+05,wkf+07}. However, in SGR~1806$-$20, there was no
evidence for a correlation between the phase-averaged flux, which
likely peaked around 2004 October, and pulsed fraction, which was
stable in the flux-enhanced pre-flare period but was lower than
average in early 2005 when the persistent flux had approached
quiescent levels \citep{rtm+05}.  Furthermore, SGR~1806$-$20's pulse
profile was noticeably more sinusoidal while brightest \citep{wkf+07},
although the pulse profile likely changed as a direct results of the
giant flare, rather than the slow evolution of persistent flux.

\section{Conclusions}

We have considered $\sim$10~yrs of multiwavelength observations
of \psr.  The source, in 2004-2006, gradually relaxed apparently
back to quiescence in most respects, only to ``awaken'' suddenly in the
glitch/flare event of 2007 March.  By observing \psr\ while quiescent,
it has become apparent that previous instabilities in timing, X-ray and
near-IR flux were likely all linked to the major long-term flaring
events of 2001-2002, whose nature is yet unknown but likely includes
changes in the stellar magnetosphere's current configuration.  The
asynchronous spin and X-ray flux variability we have observed is
incompatible with expectations of the fossil disk accretion model.
Following the most recent 2007 event, we observed total X-ray flux
variability that is strongly correlated with X-ray pulsed fraction,
X-ray spectral hardness, and changes in near-IR flux. These
observations largely agree with expectations of the magnetar model. To
date, the \psr\ 2007 event is ongoing, as is continued multiwavelength
monitoring.

\acknowledgements We thank Joe Hill and Lorella Angelini for their
help with the \swift\ XRT data analysis, and Zhongxiang Wang for his
help with the near-IR analysis. We thank the anonymous referee for
many judicious comments that have improved the quality of the
paper. FPG is supported by the NASA Postdoctoral Program administered
by Oak Ridge Associated Universities at NASA Goddard Space Flight
Center. PMW gratefully acknowledges support for this work from
NASA/SAO through grant GO7-8077A.  This research has made use of data
obtained through the High Energy Astrophysics Science Archive Research
Center Online Service, provided by the NASA/Goddard Space Flight
Center, and is based on observations made with the NASA/ESA Hubble
Space Telescope, obtained at the Space Telescope Science Institute,
which is operated by the Association of Universities for Research in
Astronomy, Inc., under NASA contract NAS 5-26555. These observations
are associated with program \#10761. This work has been supported by
SAO grant GO7-8077Z, an NSERC Discovery Grant, the Canadian Institute
for Advanced Research, and Le Fonds Qu\'eb\'ecois de la Recherche sur
la Nature et les Technologies.

%\bibliographystyle{apj}
%\bibliography{myrefs,journals1,modrefs,psrrefs,crossrefs}

\end{document}